# User-Centric Design of UI for Mobile Banking Apps: Improving UI and Features for Better Customer Experience

Luniva Chitrakar[a], Ishan Panta[a], Biplov Paneru[c], Sangharsh Poudel[a], Lahana Kansakar[a]

[a]Department of Computer Science and Engineering, Kathmandu University
[c]Independent researcher, Nepal
**corresponding author email:** thebplvstar001@gmail.com

*Abstract*

Financial management has been revolutionized by mobile banking, but increasing usefulness and satisfaction requires a better user experience. This study aims to provide an improved customer experience by offering user-friendly interfaces, and real-time notifications by user-centric design of mobile banking application UI. A survey was carried out on the target audience in which 81% of respondents to a study of 103 people said they regularly used mobile banking apps, while 77% said they had problems with the ones they were using at the time. Furthermore, 44.7% of respondents expressed unhappiness with the current solutions by depending on third-party apps like e-Sewa and Khalti for everyday transactions. Language obstacles, lengthy loading times, unclear terminology, and navigational challenges were among the problems found. With 84% asking for a budgeting function and 46% complaining about biometric authentication, users indicated a need for more individualized interfaces, improved customer service, and increased security. The study included Think Aloud testing, heat maps, and remote usability testing to determine user preferences and pain spots to solve these. Feedback from a wider audience was obtained informally through guerrilla usability testing. The results highlight how important it is for mobile banking apps to guarantee security, increase functionality, simplify navigation, and improve visual design. App grouping and layout can be further enhanced by utilizing Gestalt psychology concepts like closeness and symmetry. The goal of these user-centered insights is to promote greater happiness and adoption of mobile banking.

# 1. Introduction

Mobile banking applications have transformed the banking industry, offering customers convenient access to financial services through smartphones and tablets. As technology advances, the focus has shifted towards effectively designing software applications designing mobile interactions in the most effective manner. Human-Computer Interaction (HCI) principles play a crucial role in creating a user-centered mobile banking experience. With around 7 billion smartphone users worldwide [1], including diverse user groups such as children, the elderly, and people with disabilities, HCI directly impacts their interaction with banking apps. Our research aims to design an HCI-driven mobile banking application that addresses challenges related to navigation complexity, UI consistency, personalization, and security. By prioritizing these factors, banks can enhance user satisfaction, engagement, and loyalty. In Nepal, Kumari Bank introduced mobile banking in 2004 [2], with subsequent efforts to improve the User Interface (UI). However, HCI implementation in banking apps has been overlooked. Recognizing this gap, our study explores existing banking applications and develops a prototype that incorporates HCI principles to create an optimized banking experience.

## 1.1 Design and Development Objectives

The major objectives behind creating a mobile banking application using HCI is:
- The convenience of accessing their financial services anytime and anywhere through their smartphones or tablets.
- Accessibility for individuals in rural areas or regions with limited banking infrastructure.
- To provide an improved customer experience by offering user-friendly interfaces, and real-time notifications.
- Accessibility for blind users to use a banking application and perform transactions.

## 1.2. Problem statement

**Statement:** Despite the abundance of mobile banking applications, the pressing issue lies in the absence of Human-Computer Interaction (HCI) principles in their design and development. This deficiency poses significant challenges and drawbacks for users, as HCI principles play a crucial role in ensuring usability, user experience, and intuitive interface design. The lack of HCI implementation in mobile banking applications gives rise to complexities, poor usability, limited accessibility, and diminished user satisfaction.

**1.3. Research Question**

What are the causes of inefficiency in user experience within mobile banking apps and how can the development of a user-centric mobile app improve UI and features to enhance user satisfaction and engagement?

**1.4. Research objectives:**

The primary objectives of this research are as follows:
i) To identify and understand the common challenges and pain points faced by users of mobile banking apps.
ii) To explore practical recommendations for improving the user interface (UI) and user experience (UX) of mobile banking apps.
iii) To investigate the integration of innovative features, such as biometric authentication and personalized financial insights, to enhance the functionality and usability of mobile banking apps.
iv) To provide valuable insights and recommendations for banks and mobile application developers to create user-centric mobile banking apps that meet the needs and preferences of users.

**1.5. Literature review**

The literature review section of this study provides a comprehensive overview of existing research and scholarly works relevant to the field of mobile banking app design and user experience. This section serves as a foundation for understanding the current state of knowledge, identifying gaps and limitations in the existing literature, and informing the design and development of our own mobile banking app.

A significant study conducted in 2022 by Kaewkitipong, Chen, Han, and Ractham [3] sheds light on the projected growth of the global mobile payment market, anticipated to reach an impressive annual growth rate of 29. However, despite this remarkable growth potential, mobile banking services continue to grapple with challenges related to user engagement, sustained usage, and overall user satisfaction. This highlights the need for further exploration and improvement in the realm of mobile banking to ensure a seamless and satisfying user experience.

Jie Yu and Chompu Nuangjamnong [4] focused on the impact of transaction speed, accessibility, affordability, adaptability, ease of use, and relative advantage on customer satisfaction. The findings highlighted the importance of these factors in shaping users' perceptions and overall satisfaction with mobile banking services. By understanding and prioritizing these key factors, banks and mobile application developers can devise strategies to enhance customer satisfaction and deliver a superior mobile banking experience.

A noteworthy study conducted by Dr. Viral Bhatt and Dixita Nagar [5] delved into the factors that influence customer satisfaction in the adoption of Mobile Banking Apps. The study extensively examined both direct and indirect effects of these influencing factors, offering valuable insights for banks and mobile application developers.

An insightful study conducted by Khalid Hamid, Muhammad Waseem Iqbal, Hafiz Abdul Basit Muhammad, Zubair Fuzail, Zahid Tabassum Ghafoor, and Sana Ahmad [6] shed light on the importance of analyzing usability issues in mobile banking apps based on factors such as age, gender, exchanging partners, and user experience. The study emphasized the need to assess the effectiveness, efficiency, trustfulness, learnability, memorability, and satisfaction of mobile banking apps.

This research study examines the intricacies of designing user interfaces (UI) and user experiences (UX), as well as the roles they play, locations where they intersect, and the developing field of customer experience. Designing creative and user-focused digital goods in the digital age, where technology is advancing quickly, necessitates an awareness of the dynamic interaction between user interface and user experience. This study also looks at how general customer satisfaction will be impacted by new developments in the UI/UX space. Furthermore, this article explores the usage of artificial intelligence (AI) in the fields of user experience (UX) and human-computer interface (HCI), as well as new developments in these areas [7].

In conclusion, the literature review conducted in this study has provided valuable insights into the current landscape of mobile banking services and the factors influencing user satisfaction. The findings emphasize the significant growth potential of the global mobile payment market, while also highlighting the persistent challenges faced by mobile banking services in terms of user engagement, sustained usage, and overall satisfaction. Further research in this field holds immense potential to uncover additional factors and strategies that can further enhance the usability and reliability of mobile banking platforms. By delving deeper into user preferences, technological advancements, and emerging trends, future studies can contribute to the widespread acceptance and increased usage of mobile banking services.

Drawing upon the invaluable insights gleaned from prior research, the paper's objective is to develop a mobile banking app that goes beyond mere functional requirements, aiming to leave a lasting impression on users.

## 2. Methodology

In pursuit of our research objectives, we have employed a diverse range of methodologies to validate the prevailing issues in the field of mobile banking and to accomplish our mission of creating a user-centric mobile banking application. These methodologies have enabled us to gather comprehensive insights and ensure that our app addresses the identified challenges effectively.

### 2.1. Validating Problem Statement

A Needs Assessment survey [7] was conducted to gain insights into the user experience of current mobile banking apps, with a focus on identifying the challenges faced by users in carrying out banking tasks and their wants and needs for improved functionality and features. The survey aimed to gather data on the user's perception of mobile banking apps, their level of satisfaction with the current user experience, and their expectations for future improvements.

### 2.2. Design Process

User-centered design (UCD) [8] theory is a design approach that prioritizes user needs and requirements throughout the design process. According to UCD, the design of a product should be based on an understanding of the user's goals, tasks, and context of use. In the context of mobile banking apps, UCD helped us create apps that meet the needs and preferences of users, resulting in a better user experience and increased customer satisfaction.

### 2.3. UI Design Approach

To embrace the concept of skeuomorphism and adhere to standard design practices, our research paper employed the powerful collaborative interface design tool, Figma. With a focus on creating an optimal user interface and user experience for our mobile banking app, we implemented Schneiderman's eight golden principles, with particular emphasis on providing informative feedback, enabling easy reversal of actions, and maintaining consistency in design. The different phases of design process is depicted in figure 1.

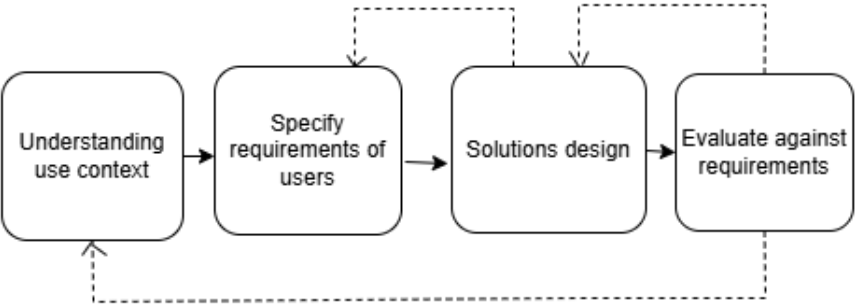

Fig 1: Phases of User Centric Design Process

Technology Acceptance Model: The Technology Acceptance Model (TAM) (Okstate Library, n.d.)is a widely used model that explains how users adopt and use technology. TAM proposes that the perceived usefulness and ease of use of a technology are key determinants of its adoption and continued use. In the context of mobile banking apps, TAM can help designers understand what features and functionalities users find useful and easy to use, which can inform the design of user-centric apps.

User Experience Design: User experience (UX) design is the process of designing products or services that provide a positive user experience. UX design focuses on understanding the user's needs, goals, and behaviors, and designing products that are easy to use, visually appealing, and meet the user's expectations. In the context of mobile banking apps, UX design can help designers create apps that are intuitive, efficient, and effective.

In line with the objective of reducing cognitive load, we implemented Miller's law [9], which limits the amount of information presented to the user at a given time. By breaking down complex tasks into smaller, manageable steps and providing clear and concise instructions or labels for each step, we aimed to enhance usability and user satisfaction.

Furthermore, we incorporated Fitt's law [10], which states that the time required to point at a target is influenced by its distance and size. By leveraging Fitt's law, we optimized the placement and size of buttons and other interactive elements within our mobile banking app, facilitating efficient and accurate user interactions.

In addition, we employed Gestalt's principles, specifically the law of proximity and the law of symmetry, to guide our design decisions [11]. These principles helped us create visually cohesive and intuitive interfaces by grouping related elements and ensuring balance and harmony in the overall layout.

### 2.4 Evaluation
#### Jacob Neilson's 10 Heuristic Evaluation

In order to evaluate our first prototype and MVP we asked a group of our classmates to perform a heuristic evaluation based on Jacob Neilson's 10 Heuristic Evaluation [12]. The goal of this exercise was to identify heuristic problems in the design of the interface and figure out how the paper could create a user interface that would eventually pass all the metrics of Jakob Nielsen's heuristic principles

#### Heat maps

To gain valuable insights into user interaction and behavior, we enlisted the participation of a select group of our classmates to test and explore our app. Throughout the testing process, the paper carefully tracked their interactions and behaviors, capturing their actions and movements within the app.

### 2.5. Testing

During the testing stage of the process, the paper used the following testing methods:

**Remote Think aloud testing**

Initially, the paper performed remote usability testing with a selected wider audience who may not have been able to participate in on-site testing. The participants were shared the app prototype and explore and provide think-aloud review. Then their interaction with the app was observed in real time through which valuable insights into participants' thought, action, and decision-making processes were gathered.

**Guerilla Testing**

Guerrilla testing, in the context of the paper, involved approaching individuals in public spaces like KU's famous local mart and food stations, explaining to them the purpose of the test and asking them to provide feedback on our mobile banking app. This approach allowed us to gather valuable insights, identify usability issues, and understand how users interacted with the app in real-world scenarios from diverse range of users who may not have prior knowledge or experience with the app.

**Comparative Analysis**

In the comparative usability testing, participants were asked to perform specific tasks using their current mobile banking app and then the paper's app. Then their interactions, challenges faced and overall experience was observed and noted.

**Heuristic Evaluation:**

After the creation of a UI, our team along with our group of peers conducted a heuristic evaluation of this user interface. For this evaluation we asked our classmates to perform heuristics evaluation of our app.

Upon receiving informative and meaningful feedback from them, we applied Jakob Nielsen's heuristic principles in order to further improve the UI and improve the mistakes we had made. The goal of this exercise was to identify heuristic problems in the design of the interface and figure out how we could create a user interface that would eventually pass all the metrics of Jakob Nielsen's heuristic principles. The heuristic evaluation based on Jakob Nielsen's heuristic are given below:

By keeping a clear view of system status and making sure users are always aware of ongoing actions, our system is built with the user experience as its top priority. For example, during the registration or form-filling processes, real-time feedback is provided by messages like "Loading" or "System error," and unambiguous indications show the number of steps done and those still to be completed. The system improves user comfort and comprehension by utilizing well-known language and design cues found in banking, such as "Are you sure?" and "Transaction successful." For improved accessibility, users can also change the interface to their favorite language. While the ability to report transactions enables users to bring issues directly to the bank's attention, features like undo actions and clear exit paths reduce errors and provide consumers a sense of control.

The application promotes a user-friendly and intuitive experience by adhering to consistent design principles throughout its menu layouts, button placements, colors, and icons. To prevent errors, we provide clear instructions and use confirmation pages to validate critical actions, along with error messages like "Field cannot be empty." By grouping operations into clearly labeled sections, providing a "Favorites" tab for commonly used features, and allowing users to search for desired functions via a search bar, the software lowers memory load. Shortcuts and customizable settings, such as the ground-breaking "Shake for Help" function, which enables users to shake their phone to report a problem or get assistance, are available to further streamline usability.

A visually calming blue color scheme has been used, information overload has been reduced by displaying only pertinent details, and succinct error messages with clear corrective instructions have been included. User confidence is guaranteed with a final transaction confirmation page, and problems are quickly reported. Users are guaranteed to have access to the aid they require thanks to extensive help resources including customer support, FAQs, and the closest bank or ATM. All things considered, our app offers a straightforward, easy-to-use, and effective banking experience.

### 2.5 Minimum Viable Product:

When creating a product, it's important to determine what the minimum set of features are that is needed to deliver to your users in order to solve their problem or provide value to them. This is what we call a minimum viable product (MVP).

It is the core feature of our mobile banking app that we have prioritized to develop first in order to quickly bring a usable product to the market. The Figma present feature serves as the foundation of our app's user interface and is essential to our user-centric approach. By focusing on this feature, we can test and gather feedback from users early on, allowing us to iterate and improve upon the app's design and functionality before adding more features. In short, the Figma present feature is the most essential and viable feature that we can develop and release to our users in order to validate our concept and move forward with the development process. It's the simplest version of your app that can still provide value to your users. This feature may not have all the bells and whistles that you want to include in the final version of your app, but it's enough to test with users and validate whether or not you're on the right track.

Now, in order to understand the users' needs, motivations, goals, and behaviors and use this knowledge to design a product that fits the users' needs and provides a better user experience, we have created the following user persona.

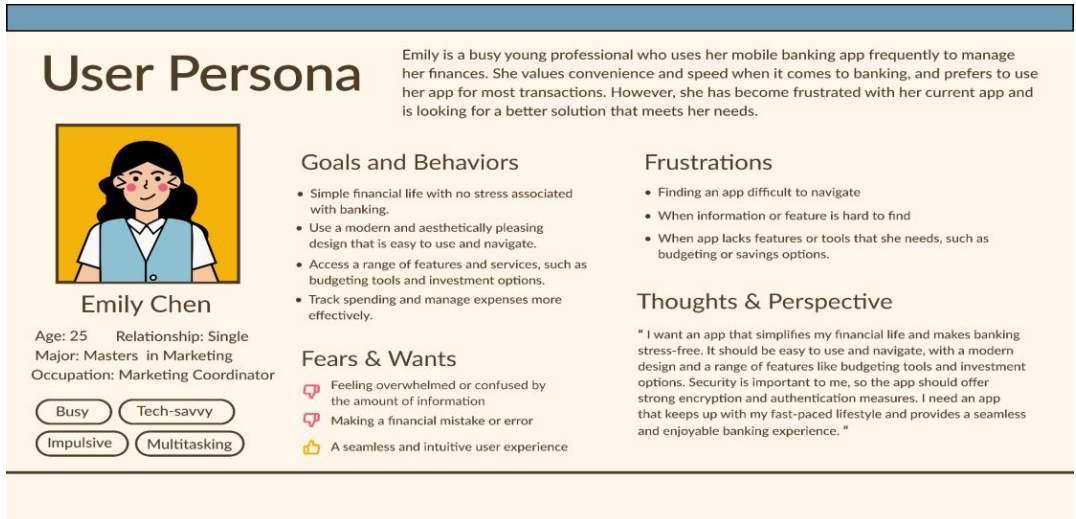

Figure 2: User Persona

Figure 2 shows the user persona. In order to understand users' needs and emotions related to our mobile banking services. It helps to create a deeper understanding of users' experiences and helps to identify areas where the product or service can be improved to better meet their needs. User perspective is depicted in figure 3.

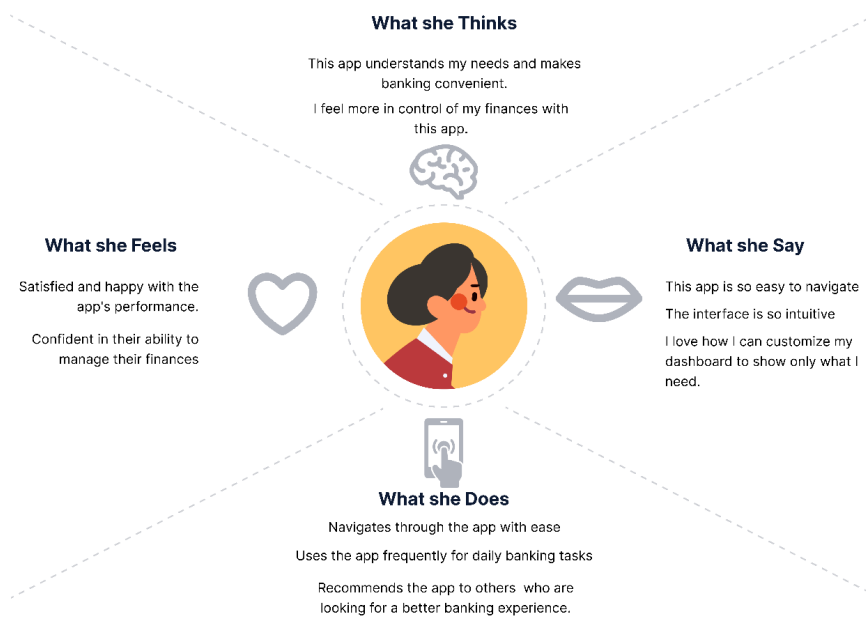

Figure 3: User perspectives

**2.6 Norman's 7 Design Principles:**

**Visibility**: We can ensure that all the available functions and options of the mobile banking app are easily visible to the user. We can achieve this by using clear and easily understandable language, icons, and menu structures.

**Feedback**: We can provide feedback to the user at every stage of the interaction with the app. For example, we can use progress bars or loading indicators to show that the app is processing the user's request.

**Constraints**: We used constraints in the design to limit the user's actions to prevent them from making mistakes. For example, we can disable buttons or options that are not relevant to the user's current task.

**Mapping**: We can use mapping to help the user understand the relationship between the controls and their effect on the system. For example, we can use clear labels and descriptions for each function and option.

**Consistency**: We can ensure that the design is consistent across the entire mobile banking app. For example, we can use the same language, colors, and design elements throughout the app.

**Affordance**: We can use affordances to provide clues as to how the mobile banking app can be used. For example, we can use buttons that look clickable or interactive.

**Discoverability**: We can ensure that all the functions and features of the mobile banking app are easily discoverable by the user. For example, we can use a clear and intuitive menu structure, and include help or tutorial features to guide the user through the app.

**2.7 Schneiderman's Eight Golden Rules**

Among the Schneiderman's eight golden rules, we have implemented the following:

**Strive for consistency**: We will aim to ensure that our app's interface is consistent across all screens and features. This means that similar elements, such as buttons and menus, should behave and look the same throughout the app. Consistent interface elements, such as buttons and menus, help users learn the app quickly and reduce cognitive load. To implement this, we can ensure that interface elements are used consistently throughout the app, and that similar actions are performed in the same way across the app.

**Offer informative feedback:** Our app will provide informative and real-time feedback to users, letting them know what is happening and what they can do to resolve any issues they encounter.

**Design dialogues to yield closure**: We will design our app's dialogues to be clear and concise, guiding users to complete their desired action without any confusion or ambiguity. This is achieved by the use of easy and understandable language in every conformation dialogues, clear steps in order to complete a task and proper visual cues.

**Permit easy reversal of actions:** Our app will allow users to easily undo their actions, making it easier for them to recover from any errors or accidental actions with clear "undo" and "cancel" buttons, familiar color pattern, icons and labels.

**Support internal locus of control**: We will design our app to give users a sense of control, allowing them to customize their experience and personalize the app to their liking in terms of font size, language, dashboard options, help resources like FAQs and progress tracking.

**Reduce short-term memory load:** Our app will be designed to minimize the cognitive load on users, ensuring that they can easily find what they are looking for without having to remember too many details or navigate through complex menus. This is achieved by keeping the interface clean and easy to understand using familiar and consistent design pattern, clear feedback, progressive disclosure and easy access to help and support.

**2.8 Miller's Law**

Miller's Law is a psychological principle that suggests that the average human can only keep 7 (plus or minus 2) pieces of information in their short-term memory at any given time. This means that if we want our app to be user-friendly and easy to use, we need to be mindful of how much information we present to the user at once.

To apply Miller's Law in our app, we can break down complex tasks into smaller, more manageable steps, and provide clear and concise instructions or labels for each step. We can also limit the number of options or choices presented to the user at any given time, and use familiar icons or symbols to help users understand the purpose of different features or functions. For example, while a user registers into mobile banking, only 3-4 input fields are kept in a single page and the whole registration service is separated into multiple pages with 3-1 inputs in each page. Similarly, each categories' payment only has 4 most used payment options displayed with horizontal sliders to scroll for more Additionally, we have used radio buttons or dropdown menus instead of free text fields to limit the number of options presented to the user at once.

**2.9 Von Restorff effect**

The Von Restorff effect, also known as the "isolation effect," refers to the phenomenon where an item that stands out from its surroundings is more likely to be remembered. In other words, when presented with a group of similar items, people tend to remember the item that is visually distinct or different.

We have implemented this effect in order to highlight our offers. Certain payments that have certain offers are made to stand out so people clearly see the offers.

## 2.10 Fitt's Law

Fitts' law is a human-computer interaction principle that predicts how long it will take for a user to point at a target on the screen, based on the size of the target and its distance from the user. The law states that the time it takes to point at a target is a function of the target's distance and size.

In our mobile banking app, Fitts' law can be used to optimize the placement and size of buttons and other interactive elements. For example, frequently used actions such as transferring money or paying bills should be prominently placed and easily accessible to reduce the time and effort required for users to complete these tasks. Similarly, buttons and links should be sized appropriately based on their importance and frequency of use. By following Fitts' law, we can create a more efficient and user-friendly interface that allows users to complete tasks quickly and with minimal effort.

## 2.11 Jakob's Law

Jacob's Law states that, "Users spend most of their time on other sites/apps, so they expect your site/app to work the same way." In other words, users are more likely to feel comfortable and confident using an interface that is similar to other interfaces they have used before.

To apply Jacob's Law in our mobile banking app, we could consider using familiar UI patterns and design elements that users are already accustomed to. For example, we could use standard icons for certain actions (such as a magnifying glass for search), use a consistent color scheme, and follow established navigation patterns. By doing so, users will feel more comfortable using the app and will be able to navigate it more easily, leading to a better overall user experience.

## 2.12 Gestalt's Principle

Gestalt's Principle is a set of principles in psychology that describe how humans tend to perceive visual elements as a whole rather than as individual parts. These principles include proximity, similarity, continuity, closure, and figure-ground. Among them, we have implemented law of

proximity and similarity

**Law of proximity:** It is grouping related elements together to visually organize information such as transactions or account details. In the app, this can be used to group related information together, such as we have created different categories of payment like utilities, school fees, etc. under which we have grouped related payments only.

**Law of symmetry:** It is creating a balanced and symmetrical layout, such as aligning buttons and input fields, to create a sense of harmony. In the app, this can be used to make sure that all buttons and interactive elements have a consistent visual style.

**2.13 Evaluation using heat maps**

During the evaluation phase, heat maps provide valuable visual representations of user interactions and behaviors. By tracking and visualizing clicks, or touches, heat maps helped us identify areas of interest, user attention, and potential usability issues within a user interface. Hence, to gain valuable insights into user interaction and behavior, authors enlisted the participation of a select group of our classmates to test and explore our app. Throughout the testing process, authors carefully tracked their interactions and behaviors, capturing their actions and movements within the app. By recording heat maps, we were able to visually analyze and understand the patterns of user attention, areas of interest, and potential usability issues. The figure 4 (a) and 4 (b) shows the heatmap applied UI.

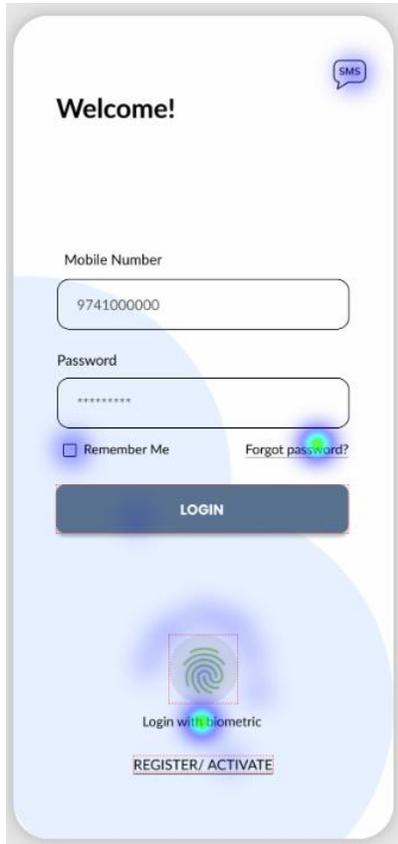
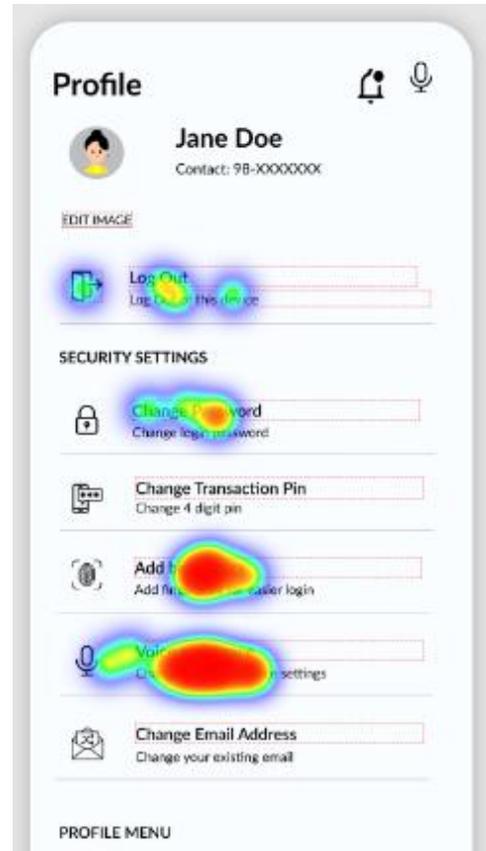

**Figure 4(a):** Login page UI with heatmap        **Figure 4(b):** Profile page UI with heatmap

## 2.14 Remote usability testing and Think Aloud testing

Initially, remote usability testing with a selected group of friends who were not physically present at Kathmandu University was performed. This approach allowed to gather feedback from a wider audience, including individuals who may not have been able to participate in on-site testing due to geographical constraints.

For remote usability testing, we utilized screen sharing and video conferencing tools to facilitate the testing sessions. The application prototype with the participants, asked the participants to shared their screens, allowing us to observe their interactions in real-time and then employed the "think aloud" method to gather valuable insights into participants' thoughts, actions, and decision-making processes as they interacted with our app. Participants were encouraged to think aloud and provide commentary on their experience, including their expectations, frustrations, confusion, and satisfaction. By asking participants to articulate their thoughts, we gained deeper insights into their cognitive processes, allowing us to understand their perspective, uncover potential usability issues and get immediate feedback. We also provided them with specific tasks or scenarios to complete while observing their interactions remotely.

Remote usability testing offered several advantages, including the ability to involve participants from different locations, accommodate their schedules, and obtain diverse perspectives. It allowed us to gather valuable feedback from a broader user base, ensuring that the app's usability and user experience considerations were not limited to a specific geographical area.

### 2.15 Guerilla Usability Testing

In this method, representative users are given specific tasks to perform using the app while their interactions and feedback are observed. Guerrilla testing, in the context of our project, involved approaching individuals in public spaces like KU's famous local mart and food stations, explaining them the purpose of the test and asking them to provide feedback on our mobile banking app. The purpose of this approach was to gather quick and informal feedback from a diverse range of users who may not have prior knowledge or experience with our app.

Participants were encouraged to think aloud and share their impressions, frustrations, and suggestions as they navigated through the app. We took notes capture important insights and observations. Unfortunately, we could record the session as the places were too crowded and loud After each session, we also asked participants specific questions to gather additional feedback on their experience.

By conducting guerrilla testing in public spaces, we were able to engage with a diverse group of people who represented one of our target user bases. This approach allowed us to gather valuable insights, identify usability issues, and understand how users interacted with the app in real-world scenarios.

## 3. Results and Analysis

In order to validate our problem statement and find out what exactly is the problem with the current mobile banking app and what do people want, we conducted a need assessment survey. The survey was highly successful, with a total of 104 insightful responses received. The significant number of participants indicated a strong level of engagement and interest in the research topic. Moreover, the survey gained a remarkable diversity of participants in relation to age, occupation, and field of expertise. The respondents represented a wide spectrum of age groups, ranging from young adults in their early 20s to middle-aged individuals in their 40s and 50s, with people from 17 different fields and 8 different occupations. This diversity proved to be highly advantageous as it allowed us to gather a wide range of opinions and perspectives.

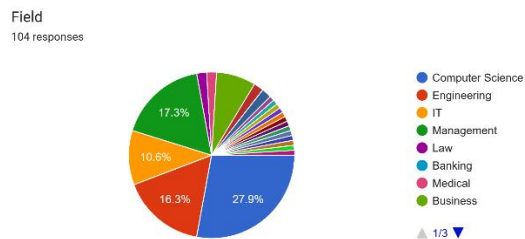

Fig 5: Diversity in user survey in terms of field

As seen in figure 5, among the 103 participants, it revealed that a significant portion (81%) of them used mobile banking apps on a daily or frequent basis. Despite the widespread familiarity and frequent usage of mobile banking apps among a significant number of people, the survey results shed light on the persistent challenges that users encounter during their interactions with these apps. Surprisingly, 77% of the participants reported facing difficulties while using their current mobile banking app.

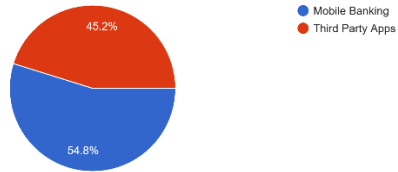

Fig 6: Comparison of user's preference between mobile banking and third-party payment app

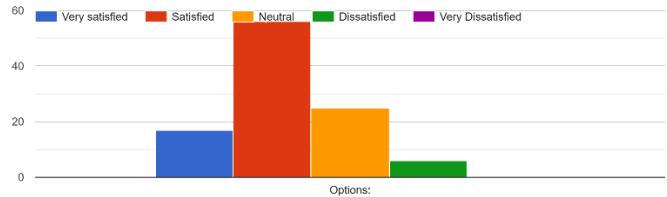

Fig 7: User satisfaction with their mobile banking experience

As depicted in figure 6 and figure 7 furthermore, 44.7% of the participants resorted to using third-party apps like e-sewa and Khalti for their day-to-day transactions, indicating a need for improvement in existing mobile banking apps.

In order to identify those usability issues, the survey also sought to identify the reasons behind the difficulties encountered by users and their expectations from an ideal mobile app. According to the survey, users faced several challenges that hinder their ability to perform banking transactions effectively and efficiently. These challenges include difficulty navigating through the app, confusing terminology, long loading times, and limited functionality, language barrier. Furthermore, users expressed a desire for more personalized and intuitive interfaces, greater transparency, and easier access to customer support. Among the respondents, 46% expressed issues with the current biometrics authentication system and expressed a desire for improvement. Additionally, 84% of the participants expressed a need for a budgeting feature in their mobile banking app.

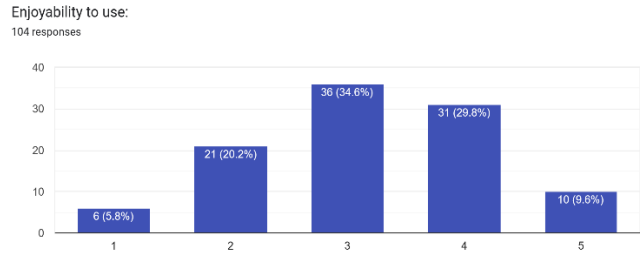

Fig 8: User rating in terms of their current mobile banking experience

Additionally, the user ratings section provided valuable insights into the overall user experience of existing mobile banking apps, covering factors such as appearance, convenience, enjoyability, consistency, feedback, short-term memory load, easy reversal, and sense of ownership. The average ratings received in these areas indicated room for improvement. The user ratings can be visualized through chart in figure 8.

In addition to gathering insights on the user experience and ratings of current mobile banking apps, we also sought input from users regarding their vision of an ideal mobile banking app. Therefore, the survey findings served as a validation and confirmation of the existence of our identified problem statement in real-life scenarios. The results strongly indicated the need for taking decisive measures to enhance the user experience in the realm of mobile banking. These findings provide compelling evidence and solidify the case for prioritizing user-centric improvements in the mobile banking sector.

## 4. Discussion

The study findings highlight the prominent challenges related to reliability and usability that hinder the adoption and effectiveness of mobile banking platforms in Nepal. Participants reported various issues and difficulties with complex user interfaces, negatively impacting their satisfaction, trust, and willingness to adopt digital payment platforms. To overcome these challenges, the paper suggests several potential strategies in order to simplifying user interfaces for enhanced usability by implementing these strategies, mobile banking platforms in Nepal can enhance their reliability, usability, and overall user experience, leading to increased adoption rates and improved customer satisfaction.

After problem validation from user survey, discussing of pros and cons and probable solutions to the existing HCI problems, the first step towards designing user centric app was paper prototyping.

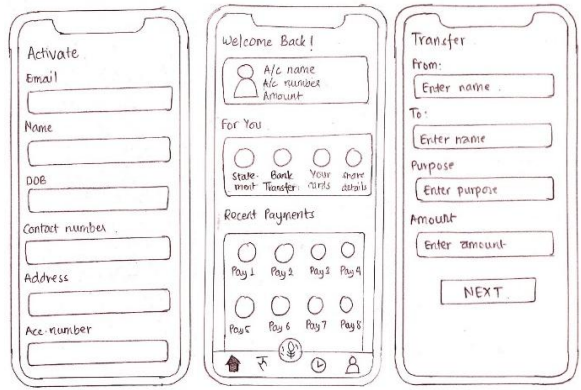

Fig 9: Initial Paper Prototype using Schneiderman's 8 golden rules

(a): UI layout I  9 (b) UI Layout II  (c) UI layout III

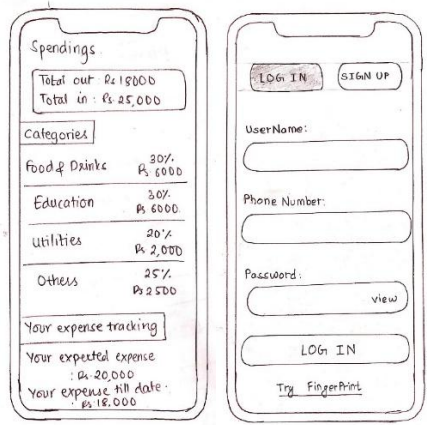

(d) UI layout IV  (e) UI layout V

The figure presents the initial prototype of our mobile banking app, adhering to the principles of Schneiderman's 8 golden rules. Figure 9 (a) showcases a simple and intuitive login page with clearly labeled input fields. In Figure 9(b), users can activate their mobile banking profile by entering their initial details. Figure 9 (c) displays the user dashboard, providing essential information such as user details, statements, and payments. Figure 9 (d) depicts the payment section, enabling users to input recipient, sender, and transfer amount. Lastly, Figure 9 (e) showcases the innovative budgeting feature, offering insights into user spending and cash flows. These designs embody our commitment to a user-centric and visually appealing mobile banking experience.

After conducting a heuristics evaluation using Jacob Neilson's principle and gathering feedback from our friends in our initial prototype, we made significant improvements to the initial paper prototype, incorporating principles such as Miller's Law, Gestalt's Law, Jacob's Law, Von Restorff effect and Fitt's Law.

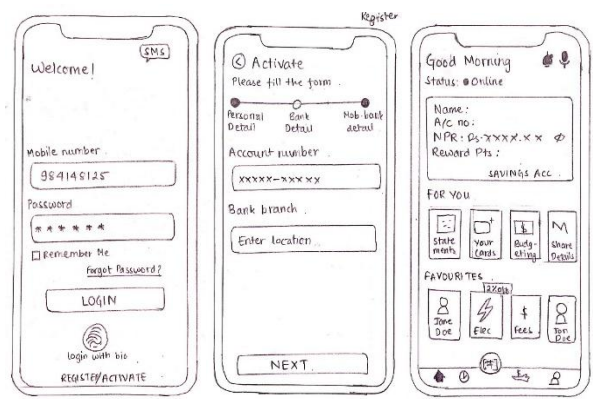

Figure 10(a): Improved login page Figure 10(b): UI registration page Figure 10(c): Streamlined dashboard

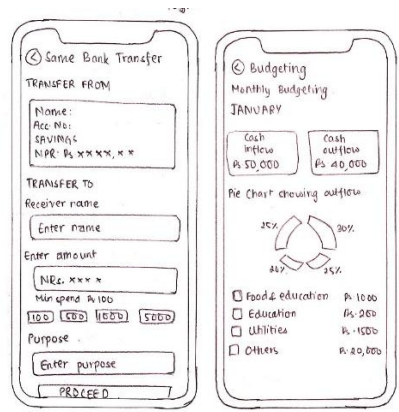

Figure 10(d): Payment page                Figure 10 (e)The budgeting feature

Fig 6: Final Paper Prototype after Heuristic Evaluation

Figure 11 showcases the revised and enhanced version of the paper prototype, focusing on user experience and visual aesthetics. Figure 10(a) highlights the improved login page, featuring a clean and uncluttered layout with enhanced emphasis on biometrics for easy login. Figure 10(b) presents a more user-friendly registration page, with categorized information and a horizontal slider indicating progress. Figure 10(c) showcases a streamlined dashboard, displaying only essential details in a clear format, along with an improved bottom navigation bar. In Figure 10(d), the payment page has been refined to provide users with relevant details, minimizing the chances of entering incorrect information. Additionally, an easy-to-use amount input section has been implemented. Finally, Figure 11(e) illustrates the budgeting feature in a visually appealing and intuitive format, designed to enhance user enjoyment and engagement.

The utilization of paper prototyping allowed us to create low-fidelity prototypes of the mobile banking application. This approach proved to be highly valuable in the early stages of the design process, as it facilitated quick iterations and feedback gathering.

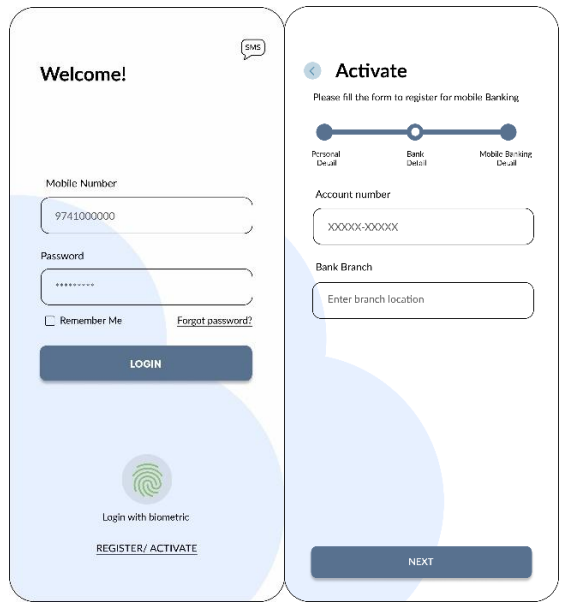

Figure 11(a): Main login page     Figure 11 (b): Verification UI

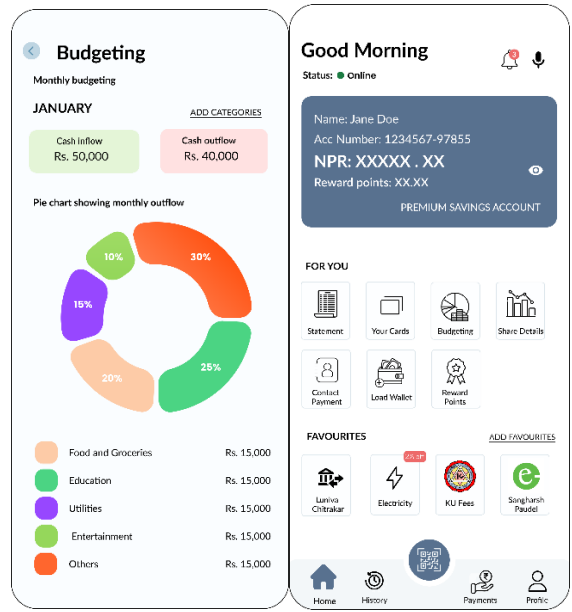

Figure 11 (c): Budgeting UI dashboard  Figure 11 (d): Final UI page

The UI design phase played a crucial role in enhancing the visual appeal and usability of the mobile banking application. We incorporated principles of user-centered design, emphasizing clarity, consistency, and intuitiveness. The final UI design as seen in figure 11 (a), 11 (b), 11 (c), 11 (d) aimed to streamline navigation, improve information hierarchy, and ensure a cohesive and visually pleasing user experience. Through user centric design processes and user feedback, we iteratively refined the UI design to meet user expectations and preferences.

To assess the usability of the mobile banking application, we conducted a series of usability tests. The tests included Remote usability testing and Think Aloud testing, Guerilla Usability Testing and Comparative Usability Testing. These tests involved a diverse group of participants, representative of the target user base. The tests allowed us to observe and gather insights into how users interacted with the application, identified usability issues, and measured user satisfaction. The usability tests revealed valuable information about areas for improvement, such as transaction flow, navigation efficiency, and clarity of instructions.

Overall, the findings from the usability tests highlighted the effectiveness of the applied design strategies and enhancements. Users reported improved satisfaction, enhanced task completion rates, and a reduction in errors. The usability tests provided empirical evidence of the positive impact of the user-centric approach, paper prototyping, and UI design on the overall user experience.

**Comparative Usability Testing:**

In the comparative usability testing, we asked participants to perform specific tasks using their current mobile banking app. We observed and noted their interactions, challenges faced, and overall experience during this task.

Following that, we introduced our mobile banking app and asked participants to explore its features and perform the same tasks. We encouraged them to provide feedback on the strengths and weaknesses they observed compared to their current app.

By conducting this comparative testing, we were able to gain insights into how our app stacked up against existing solutions in terms of usability and user experience. Participants' feedback on the pros and cons of our app helped us identify areas of improvement and validate the effectiveness of our user-centric design approach. The pros and cons that we got from this comparative testing are tabulated in table 1.

**Table 1:** UI design prospects

| Aspect | Details |
|---|---|
| **Pros** | |
| Intuitive Interface | Participants found the app to have a user-friendly interface, making navigation and task performance easier. |
| Simplified Processes | The app streamlined complex banking processes, making transactions and financial management more convenient and efficient. |
| Customization Options | Offered features like setting transaction limits and notification preferences for a personalized experience. |
| Helpful Error Messages | Provided informative and user-friendly error messages to assist users in resolving issues or correcting input errors. |
| **Cons** | |

| Limited Features | Lacked certain advanced and necessary features, such as additional payment demonstrations. |
|---|---|
| Learning Curve | Some users experienced a slight learning curve due to new features or different workflows compared to their current app. |
| Prototype Limitation | Users reported a lack of tactile and immersive feeling as it was a Figma prototype and not fully functional. |

Analyzing the heatmap data, we observed that the login page effectively utilized essential components, ensuring a streamlined user experience. Each component seems to be essential and not unnecessary. The inclusion of biometrics as a login option proved beneficial, providing users with a convenient and secure method to access their mobile banking account. Furthermore, we noticed that users showed a tendency to explore the SMS mode while offline. However, we realized that we had not developed that feature to its full potential, highlighting the need for further improvement. This valuable insight prompted us to recognize the importance of incorporating a well-developed SMS mode into our mobile banking app.

We also noticed a prominent trend of users relying heavily on the bottom navigation bar to navigate between different pages, each serving distinct functionalities. This feature proved to be intuitive and facilitated seamless transitions for users. Additionally, the frequent utilization of the "back" button indicated that it effectively enabled users to easily reverse their actions and navigate back to the previous page.

It was interesting to note that the voice assistance feature garnered significant user engagement. This can be attributed to the novelty and curiosity surrounding voice-based interactions in the context of mobile banking. The popularity of this feature highlights its potential to enhance the user experience and should be further explored and optimized.

Overall, the cognitive walkthrough revealed positive user interactions with the login page, navigation bar, "back" button, and voice assistance feature. These findings validate the effectiveness of the design choices made and serve as valuable insights for further improving the user experience of our mobile banking app.

**Limitations and Future Scopes**

In this paper, the focus on a user-centric approach and specific tools and techniques of HCI resulted in not incorporating all available features of the existing mobile banking application. Due to new features or different workflows than their current software, some customers encountered a slight learning curve. While this may have overlooked some potentially valuable features, the selected enhancements were based on user requirements and aimed at improving the overall user experience. Moreover, the use of a Figma prototype limited the tactile and immersive experience for users, which could have affected their perception and evaluation of the application. Future research could explore additional features and incorporate more interactive simulations to provide a more realistic user experience during testing and evaluation.

## 5. Conclusion

In conclusion, this work emphasized the importance of incorporating Human-Computer Interaction (HCI) principles in remodeling a mobile banking application to enhance user experience. Through a comprehensive analysis that involved interviews, surveys, prototyping, UI development, and evaluation procedures, we gained valuable insights into user requirements and preferences. The findings highlight the significance of reducing cognitive load and creating a user-centric design. By implementing these strategies, we aim to develop a mobile banking application that meets user needs and enhances overall usability and satisfaction.

## 6. References

[1] Kaewkitipong, L., Chen, C., Han, J., & Ractham, P. (2022). Human–Computer Interaction (HCI) and Trust Factors for the Continuance Intention of Mobile Payment Services. *Sustainability*.

[2] Bhatt, D. N. V. (2021). An Empirical Study to Evaluate Factors Affecting Customer Satisfaction on the Adoption of Mobile Banking Services. *Turkish Journal of Computer and Mathematics Education*.

[3] BankMyCell. (2023, May). How many smartphones are in the world? [Online]. Available: https://www.bankmycell.com/blog/how-many-phones-are-in-the-world

[4] Nepal Journals Online. (n.d.). Opportunities and Challenges in Mobile Banking in Nepal.

[5] Kaewkitipong, L., Chen, C., Han, J., & Ractham, P. (2022, September). Human–Computer Interaction (HCI) and Trust Factors for the Continuance Intention of Mobile Payment Services. *Sustainability*.

[6] Yan, J., & Nuangjamnong, C. (2022, August). The Impact of Mobile Banking Service on Customer Satisfaction: A Case Study of Commercial Banks in China. [Online]. Available: https://www.researchgate.net/publication/363007742_The_Impact_of_Mobile_Banking_Service_on_Customer_Satisfaction_A_Case_Study_of_Commercial_Banks_in_China

[7] Nagar, D., & Bhatt, V. (2022). An Empirical Study on Customer Satisfaction of Mobile Banking Services. [Online].

[8] Hamid, K., Iqbal, M. W., Basit, H. A., Fuzail, Z., Ghafoor, Z. T., & Ahmad, S. (2022). Usability Evaluation of Mobile Banking Applications in the Digital Era. [Online].

[9] Community Tool Box. (2022). Conducting Needs Assessment Surveys. [Online]. Available: https://ctb.ku.edu/en/table-of-contents/assessment/assessing-community-needs-and-resources/conducting-needs-assessment-surveys/main

[10] Interaction Design Foundation. (n.d.). User Centered Design. [Online]. Available: https://www.interaction-design.org/literature/topics/user-centered-design

[11] Greenan, R. (2023, March 2). What is Miller's Law in UX Design? A Complete Guide. [Online]. Available: https://careerfoundry.com/en/blog/ux-design/what-is-millers-law/

[12] Interaction Design Foundation. (2022). Fitts' Law. [Online]. Available: https://www.interaction-design.org/literature/topics/fitts-law

[13] Interaction Design Foundation. (n.d.). The Gestalt Principles. [Online]. Available: https://www.interaction-design.org/literature/topics/gestalt-principles

[14] Nielsen, J. (2020). 10 Usability Heuristics for User Interface Design. [Online]. Available: https://www.nngroup.com/articles/ten-usability-heuristics/

[15]     Paneru, B., Paneru, B., Poudyal, R., & Shah, K. B. (2024, March). Exploring the Nexus of User Interface (UI) and User Experience (UX) in the Context of Emerging Trends and Customer Experience, Human Computer Interaction, Applications of Artificial Intelligence. *INJIISCOM, 5*(1), 102-113.
3